\documentstyle[prd,aps,floats]{revtex}
\def\spose#1{\hbox to 0pt{#1\hss}}
\def\lta{\mathrel{\spose{\lower 3pt\hbox{$\mathchar"218$}}
     \raise 2.0pt\hbox{$\mathchar"13C$}}}
\def\gta{\mathrel{\spose{\lower 3pt\hbox{$\mathchar"218$}}
    \raise 2.0pt\hbox{$\mathchar"13E$}}}

 \def\mP{m_{_{\rm P}}}       \def\mp{m_{\rm p}}

 \def\rhv{\rho_{\rm v}}

 \def\U{{\cal U}}                    \def\Th{\Theta}

\begin{document}
\preprint{BROWN-HET-...., hep-ph/yymmddd}
\draft

\newcommand{\vp}{\varphi}
\newcommand{\be}{\begin{equation}} \newcommand{\ee}{\end{equation}}
\newcommand{\bea}{\begin{eqnarray}} \newcommand{\eea}{\end{eqnarray}}

\renewcommand{\topfraction}{0.99}
\renewcommand{\bottomfraction}{0.99}
\twocolumn[\hsize\textwidth\columnwidth\hsize\csname 
@twocolumnfalse\endcsname

\title{Microwave Background Constraints on Decaying Defects}

\author{Robert H. Brandenberger$^{1,2}$, Brandon Carter$^3$  and 
Anne-Christine Davis$^4$}

\bigskip

\address{$^1$ TH Division, CERN, 1211 Geneva 23, Switzerland (address from
9/15/01 - 3/15/02);\\
$^2$ Physics Department, Brown University, Providence, R.I. 02912, USA,\\
e-mail: rhb@het.brown.edu;\\
$^3$ LUTH, Observatoire de Paris-Meudon, 92 195 Meudon, France,\\
e-mail: carter@obspm.fr;\\
$^4$ DAMTP, Center for Mathematical Sciences, University of Cambridge,\\
Wilberforce Road, Cambridge, CB3 0WA, U.K.,\\
e-mail: A.C.Davis@damtp.cam.ac.uk .}

\date{15 February 2002}

\maketitle

%{\bf Abstract}
\begin{abstract}
Embedded defects are predicted in a host of particle physics theories,
in particular in the standard electroweak theory. They can be 
stabilized by interactions with the cosmological plasma, but will decay 
once the plasma falls out of equilibrium, emitting a substantial fraction 
of non-thermal photons. If the decay happens after a redshift of about 
$10^6$, these photons will give rise to spectral distortions of the 
Cosmic Microwave Background, which leads to strong constraints on the 
underlying particle physics theory. Such constraints apply to any model 
with decaying defects, and in particular to theories predicting decaying 
vortons, thereby leading to constraints stronger than the dark matter 
limit.

\end{abstract}

]

%\vfill\eject

%\baselineskip 24pt plus 2pt minus 2pt
%\baselineskip 15pt plus 2pt minus 2pt
\vskip 0.4cm

\section{Introduction}
\label{sec:0}

It has recently been realized \cite{Nagasawa:1999iv} that many
types of embedded defects can be stable in the cosmological
plasma of the early Universe. Particular examples are the
Z-string \cite{Vachaspati:dz} of the standard electroweak theory and the
pion strings of the Sigma model description of low energy QCD in the
limit of vanishing pion mass \cite{Zhang:1997is}, which are both
stabilized by interactions with the electromagnetic plasma. Since
the photons fall out of thermal equilibrium at the time $t_{\rm ls}$ of
last scattering, the stabilization forces will disappear at this time
and the embedded defects will decay, emitting a certain fraction of
their energy as non-thermal photons.

Cosmological models with non-thermal photon production occurring after
a redshift of $10^6$ are strongly constrained by the precision measurements
\cite{FIRAS}
of the black-body spectrum of the Cosmic Microwave Background (CMB).
The fraction of photons produced after the above redshift must be
smaller than $7 \times 10^{-5}$, otherwise the Compton y parameter
measuring the spectral distortion of the CMB would be larger than the
observational limits.

The purpose of this Letter is to study the constraints on theories
with embedded defects which result from the FIRAS limits. Our analysis
is applicable to any theory with decaying topological defects. As we
will see, the strongest constraints result for theories with vortons
\cite{Davis:ij}. The resulting constraints are in many  cases stronger 
than the nucleosynthesis and dark matter constraints on vorton models 
obtained in previous work~\cite{Brandenberger:1996zp,CarterDavis:99}.  

The outline of this Letter is as follows. We first review why some
embedded defects can be stabilized by the cosmological plasma of the
early Universe. We then focus on models with embedded strings,
such as the standard electroweak theory, or decaying
topological strings and derive the constraints
in several circumstances, depending on whether the strings have
current and form vortons or not, and depending on whether they 
decay during the string scaling regime or in the friction-dominated
regime. In the final section we discuss our results.

\section{Stabilization and Decay of Embedded Defects}
\label{sec:1}

To illustrate the mechanism that renders some embedded defects stable
in the early Universe, we shall consider a model~\cite{Carter:2002te} 
in which the order
parameter consists of four real scalar fields with a standard
symmetry breaking potential symmetric in the four fields. We will
assume that two of the fields are electrically charged and the other
two neutral, as is the case in the standard electroweak theory and
in the Sigma model description of low energy QCD in the limit of
vanishing pion mass.

In an electromagnetic plasma, the interactions with the photon bath
will lead to an effective potential for the order parameter which
breaks the symmetry between the charged and the neutral fields, while
preserving the symmetry within the neutral scalar field sector. The
potential is lifted more in the direction of the charged scalar fields.
Thus, the vacuum manifold becomes $S^1$, giving rise to cosmic
string solutions of the full field equations which look like the
standard $U(1)$ cosmic string configuration of the neutral fields
with the charged scalar fields set to zero.

The thermal averaging implicit in the above analysis breaks down
after the time of last scattering, and thus it is expected that
the confining potential will disappear and the embedded strings will
decay, emitting a fraction $f$ (which is expected to be smaller but
of the order of $1$) of its energy as photons. Since the topological
defects are out-of-equilibrium objects, the photons produced by
their decay will lead to spectral distortions of the thermal CMB
\footnote{Note that the non-thermal nature of the decay products
of strings has recently been invoked as a possible source of
non-thermally distributed cold dark matter \cite{Jeannerot:1999yn}.}

Let us now look at this stabilization mechanism in more detail.
The Higgs sector of our model is described by the following
Lagrangian
\be \label{01} 
{\cal L} \, = \, {1\over 2}
\big(\chi_{|\mu}\,^\ast\!\chi^{|\mu}+\phi_{;\mu}\,^\ast\!\phi^{;\mu} \big) 
- V\, ,
\ee
with the symmetry breaking potential
\be \label{02} 
V \, = \, {\lambda\over 4}\big(\chi\,^\ast\!\chi+ 
\phi\,^\ast\!\phi-\eta^2\big)^2  \, ,
\ee
and with gauge-covariant derivative
\be \label{03}
\chi_{|\mu} \, = \, \chi_{;\mu}+ie A_\mu\chi \, ,
\ee
where $e$ is the fundamental charge, $A_{\mu}$ is the gauge field
of electromagnetism, and a semicolon denotes the space-time
covariant derivative. The field $\phi$ denotes the neutral Higgs
doublet, the field $\chi$ the charged Higgs doublet. Using the
language of the low-energy Sigma model of QCD we can write
\be \label{04} 
\chi \, = \, \pi_1+ i\pi_2\, ,\hskip 0.7 cm \phi \, = \, \pi_3+i\pi_0
\, .
\ee 

The finite temperature corrections to the potential of the theory given
by (\ref{01}) were worked out in detail in \cite{Carter:2002te}. 
Keeping only the contributions to the finite temperature effective
potential $V_{_\Th}$ quadratic in the temperature $\Th$ we obtain
\begin{eqnarray} \label{05}  
V_{_\Th} - V \, &=& \, {{\Th^2} \over 2} e^2\Big({1\over 6}
A_\mu A^{\mu}+ \pi\, \chi\,^\ast\!\chi\Big)\nonumber\\ 
 &+& {\Th^2}{\lambda\over 4}\big(\chi\,^\ast\!\chi+
\phi\,^\ast\!\phi - {2\over 3}\eta^2) \, .
\end{eqnarray}
The second term in the first line is responsible for the breaking of
the degeneracy in the potential between the charged and neutral scalar
fields. Equation (\ref{05}) describes a potential which is lifted by
different amounts in the neutral and charged scalar field directions
compared to the zero temperature potential. For temperatures below
the critical temperature $\Th_c = \sqrt{2} \eta$, the space of lowest
energy states forms a manifold $S^1$ consisting of configurations
with $\chi = 0$ and $\phi^{\ast} \phi = \eta_{_\Th}^{\,2}$, where 
$\eta_{_\Th}$ is given by
\be \label{06} 
\eta_{_\Th}^{2} \, = \, \eta^2- {{\Th^2} \over 2}\, .
\ee 
Thus, there is a static string solution consisting of a $U(1)$ cosmic
string in the $\phi$ variables with $\chi = 0$, the {\it embedded string}.
As for usual cosmic strings, the mass per unit length $\U_0$ of this 
embedded string is given by
\be \label{07}
\U_0 \, = \, \alpha \eta^2 \, ,
\ee
where $\alpha$ is a numerical factor of order $1$.

The embedded string is stable in the temperature range immediately below
the critical temperature for which the curvature of the potential in
the $\chi$ direction is positive at $\chi = 0$. This is the case for
\be \label{08} 
\sqrt{2}\Big(1+{2\pi e^2\over\lambda}\Big)^{-1/2} \, < \, {\Th\over\eta} \,
< \, \sqrt{2}\, .
\ee

When the temperature drops below the lower limit given in (\ref{08}), 
the string does not disappear, but rather undergoes a core phase 
transition \cite{Axenides:1997ja} 
(see also \cite{Axenides:1997sk,Axenides:1998yz})
in which the charged field $\chi$ 
acquires a non-vanishing value $\chi_{_\Th}$ in
the string core which lowers the potential energy density in the core.
What results is an asymmetric vortex defect which, as 
realized in \cite{Carter:2002te}, will generically be superconducting,
the superconductivity being induced by the phase gradient of $\chi$ in
the string core. The value of $|\chi_{_\Th}|$ is given by
\be \label{09} 
\chi\,^\ast\!\chi \, = \, \eta^2-{{\Th^2} \over 2}\Big(
1+{2\pi e^2\over\lambda}\Big)\, .
\ee

The potential energy density $V_{_\Th}$ in the core of an asymmetric
vortex defect is smaller than the corresponding energy density 
$V_{_\Th, 0}$ of the symmetric embedded defect. Combining the bare 
potential (\ref{02}) with the quadratic temperature corrections given 
by (\ref{05}) we find the following expression for the finite 
temperature effective potential $V_{_\Th}$
\begin{eqnarray} \label{10} 
V_{_\Th} \, 
&=& \, e^2{{\Th^2} \over 2} \Big({1\over 6} A_\mu A^\mu+
\pi\chi\,^\ast\!\chi\Big)    \nonumber \\
&+& \, {\lambda\over 4}\big(\chi\,^\ast\!\chi+\phi\,^\ast\!\phi
-\eta_{_\Th}^{\ 2} \big)^2+C_{\Th}\, ,
\end{eqnarray}
where $C_{_\Th}$ is a constant depending on temperature.

The potential energy density $\Delta V_{_\Th}$ in the core of the
asymmetric embedded defect is obtained by evaluating (\ref{10})
at the center of the core, i.e. for $\phi = 0$ and $\chi$ given by
(\ref{09}), and subtracting the constant $C_{_\Th}$. 
Let us now consider temperatures $\Th$ much lower than the critical 
temperature $\Th_c$. In this case, the result is
\be \label{11}
\Delta V_{_\Th} \, \simeq \, {{e^2} \over 2} \pi \Th^2 \eta^2 \, .
\ee
The radius $r_{\chi}$ of the asymmetric vortex is given by equating
the potential energy density and the tension energy within the core
radius, with the result
\be \label{12}
r_{\chi} \, \sim \, {{(\chi^{\ast} \chi)^{1/2}} \over 
{\Delta V_{_\Th}^{1/2}}}\, \simeq \, \bigl({2 \over \pi}\bigr)^{1/2} 
e^{-1} \Th^{-1} \, .
\ee
For small gauge coupling constant, the width of the asymmetric vortex
thus turns out to be larger than the thermal length, thus justifying
our use of thermal averaging to study the local vortex structure.
Combining (\ref{11}) and (\ref{12}), we see that the energy per unit
length $\U$ is of the same order of magnitude as $\U_0$, namely
\be \label{13}
\U \, \sim \, \eta^2 \, .
\ee
This demonstrates that most of the energy loss of embedded defects
occurs when they finally decay.

Since the asymmetric vortices are superconducting, string loops will
generically form vortons \cite{Davis:ij}. The strings acquire their
current at the time $\Th_{_{\rm Q}}$ of the core phase transition, 
i.e. (see (\ref{08}))
\be \label{14}
\Th_{_{\rm Q}} \, = \, \sqrt{2} \Big(1+{2\pi e^2\over\lambda}\Big)^{-1/2} 
\eta \, .
\ee
Since this temperature is only slightly lower than the temperature
at which the strings initially form, the string network will still
be in the friction-dominated phase when the current condensation
occurs. This information is crucial (see \cite{Brandenberger:1996zp})
for determining the vorton density.

In the following, we will use the knowledge of the energy contained
in embedded vortices, in particular those of asymmetric core nature,
to determine the cosmological constraints.

\section{Constraints}
\label{sec:2}

Cosmic vortices (and other defects) that decay into 
(in part) photons at some 
temperature $\Th_{\rm d}$ corresponding to a redshift of 
less than $10^6$
will produce spectral distortions of the CMB and will thus be strongly
constrained by the COBE/FIRAS data~\cite{FIRAS}. The constraint
on the non-thermal fractional energy density production in photons is
\be \label{fir}
{{\delta \rho_{\gamma}} \over {\rho_{\gamma}}} \, \lta \, 7 \times 10^{-5}
\ee 
This constraint arises from the limits to the Compton $y$ parameter which
measures the spectral distortion. Any
defects which decay after a redshift of $10^6$ will be subject to these FIRAS 
constraints. Thus, the following considerations will apply to
stabilized embedded defects and to decaying topological defects (such a
decay might be induced by a late time phase transition).

The specific bounds on defect models will depend on the type of defect,
on the density of defects and on the specifics of the decay. We will
focus on cosmic strings, both of topological and of embedded type.
First, we consider strings which are in their scaling regime at the time
of decay (see e.g. \cite{ShellVil} for a review of cosmic string
dynamics). This applies, for example, to embedded vortices formed at early
times and in which superconductivity is absent or too weak to produce
vortons. It also applies to topological, non-superconducting strings
formed at early times. If the strings are formed later than some
critical time (whose value is derived below), the string network will
not yet be scaling and the dynamics will still be friction-dominated.
Since the density of strings is higher in this phase relative to a
string scaling configuration, the bounds in this case are different.
This is the second case we treat. Finally, we analyze the - from the point 
of view of embedded asymmetric vortices most realistic - case of a gas 
of vortons produced by the superconducting string loops. Cosmological 
constraints on models with vortons have previously been studied 
(see e.g. \cite{Brandenberger:1996zp,CarterDavis:99} 
and references therein). The constraints we derive here are tighter 
(but obviously only apply in the case of decaying vortons such as those 
of interest in the case of embedded defects). 

\subsection{Scaling Cosmic Strings}

First let us examine the case of cosmic strings in the scaling regime. 
Cosmic strings formed at a temperature $\Th_c$ with a mass per unit length
of $\U$, which is approximately $\Th_c^2$, will be in the 
scaling regime after a period of time $t_{\rm x}$, where \cite{Kibble:1981gv}
\be \label{16}
t_{\rm x} \, \approx \, \left(G\U\right)^{-1}t_c \, \approx \,
\bigl({{45} \over {32 \pi^3}}\bigr)^{1/2}\left(\mP\over \eta \right)^4
\mP^{-1}
\ee
where $\U\approx\eta^2$.
In the scaling regime the density of strings at time t is
\be \label{17}
\rho_{\rm s} \, = \, \nu {{\eta^2} \over {t^2}} \, ,
\ee
where the constant $\nu$ determines the number of long string segments per
Hubble volume, and whose value is $\nu \sim 10$ (see \cite{ShellVil} and
references therein).

Let us first consider strings decaying at the time $t_{\rm ls}$ of last
scattering. From (\ref{17}) it follows that the density of strings at 
last scattering is
\be \label{18}
\rho_{\rm s} \, = \, \nu {{32\pi^3}\over {45}} {{z_{\rm eq}}
\over{z_{\rm ls}}} {{\eta^2}\over{\mP^2}} \Th_{\rm ls}^{\, 4} \, ,
\ee
where $z_{\rm eq}$ is the redshift at the time of equal matter and radiation.
If a fraction $f$ of the energy of the strings goes into photons, then 
the COBE/FIRAS constraint (\ref{fir}) becomes
\be \label{19}
{{f\rho_{\rm s}}\over{\rho}} \, = \, 
f \nu {{32\pi} \over 3} \left(\eta \over \mP\right)^2 
{{z_{\rm eq}} \over {z_{\rm ls}}} \, \lta \, 7 \times 10^{-5}.
\ee
This yields a constraint on the scale of symmetry breaking
\be \label{20}
{\eta\over \mp} \, \lta \,
({7 \times 10^{-5}})^{1/2} \left({{z_{\rm ls}} \over {z_{\rm eq}}}
{3\over{32 \pi}}
\right)^{1/2} f^{-1/2} \nu^{-1/2}.
\ee
 For example, assuming that $f$ is of order $1$ then (\ref{20})
results in the constraint 
\be \label{21}
\eta \, \lta \, \nu^{-1/2} 10^{16} {\rm GeV}.
\ee
Thus, for values of $\nu$ in the range indicated by present cosmic string
simulations, the COBE/FIRAS constaint severely constrains decaying
cosmic string models with a symmetry breaking scale given by the scale of
Grand Unification. Note, however, that all cosmic string models with
this scale of symmetry breaking are already constrained 
\cite{Albrecht:2000hu,Durrer:2001cg,Magueijo:2000se} by the fact
that they would produce observable isocurvature fluctuations in the
CMB of amplitude larger than what is compatible with recent CMB
anisotropy observations. 

Consider now strings decaying at a redshift $z_{\rm d}$ larger than 
$z_{\rm eq}$ (but smaller than $10^6$). In this case, the COBE/FIRAS 
constraint yields a result analogous to (\ref{20}), but without the factor 
of $(z_{\rm ls} / z_{\rm eq})^{1/2}$ (a factor resulting from the faster
redshifting of the radiation energy density after $t_{\rm eq}$ compared to
that of the string energy density). Thus, the bound is weaker by a
factor of about 3 than the bound given in (\ref{21}). Conversely, if
the strings decay after $t_{\rm ls}$, the bound is stronger by a factor
of $(z_{\rm d} / z_{\rm ls})^{1/2}$, a fact which again follows 
immediately from the different rates of redshifting of energy densities.

\subsection{Friction Dominated Strings}

Non-conducting cosmic strings will remain in the friction dominated phase in 
which they strongly interact with the surrounding plasma until the time 
$t_{\rm x}$
given in (\ref{16}), which corresponds to the Kibble limit 
temperature given by
\be \label{22}
\Th_{\rm x} \, = \, \bigl({{32 \pi^3} \over {45}}\bigr)^{1/4}
\mP \left(\eta\over\mP\right)^2
\ee
This friction domination condition will still be satisfied at the time 
when the strings decay (assumed to be $t_{\rm ls}$) whenever $\Th_{\rm x}$ 
is less than $\Th_{\rm ls}$, i.e. less than $10^{-13}$ GeV. This gives
\be \label{23}
{\eta\over\mP} \lta 10^{-16},
\ee
i.e. $\eta \lta 10^3$ GeV. 
As well as applying to non-conducting strings, Kibble's lower bound 
(\ref{22}) is also relevant for strings with currents of the 
non-electromagnetic chiral type~\cite{CarterDavis:99}, but for electrically 
conducting strings the epoch of friction domination can be greatly 
prolonged~\cite{DimopoulosDavis:98,DimopoulosDavis:99,Carter:2000fv}, 
and in consequence the limit (\ref{23}) will be considerably relaxed.

The network of strings in the friction epoch can be described as a
Brownian walk network with scaling length given by 
\cite{ShellVil,Brandenberger:1996zp}
\be \label{24}
L\{\Th\} \, \approx \, \mP^{1/2} {{\Th_c} \over {\Th^{5/2}}} \, .
\ee
The corresponding energy density in the string network is
\be \label{25}
\rho_s\{\Th\} \, \sim \, \nu {{\U} \over {L\{\Th\}^2}} \, \sim \,
\nu \mP^{-1} \Th^5 \, .
\ee
Note that the scale of symmetry breaking $\eta$ has cancelled out.
This is due to the fact that as $\eta$ increases, the mass per unit
length of each string increases, but the string density decreases
because we are considering later times in the friction epoch.
From the $\Th^{5}$ scaling of $\rho_s$ it follows immediately that
strings in the friction epoch are not constrained by the COBE/FIRAS
data, unless they lead to vortons.

Even if they contained no currents, strings formed at the electroweak
limit would not be excluded by the limit (\ref{23}).
However, electroweak strings will have quark and lepton
zero modes in their core \cite{Davis:1996xs} and are thus current-carrying.
Current-carrying strings give rise to stationary loops, or vortons,
and will therefore be subject to the constraints discussed in the
next section.
  
\subsection{Vortons}

If any cosmic strings, either embedded strings or topologically stable
strings, are current-carrying, then string loops would form a stationary
equilibrium state or vorton\cite{Davis:ij}. For topological strings, the 
vortons could be very stable, particularly in the chiral 
case~\cite{CarterDavis:99}. In earlier work~\cite{Brandenberger:1996zp} 
we have considered constraints on particle physics theories yielding 
vortons from nucleosynthesis and dark matter considerations. If the 
vortons survive at least a few minutes, then a constraint can be derived 
by demanding that the Universe remain radiation dominated at the time of 
nucleoysynthesis. If the vortons are absolutely stable or decay with a 
lifetime greater than the age of the Universe, an additional and stronger 
constraint follows by requiring that the vorton density does not overclose 
the Universe. 

The main point of the present work is to  consider the additional 
constraint arising from the limit (\ref{fir}) provided by the COBE/FIRAS 
data~\cite{FIRAS}  in cases for which vortons decay between a redshift 
of $10^6$ and today, thus giving rise to decay products that would 
produce observable distortions of the black body spectrum.

If the string becomes current-carrying at a time $\Th_{_{\rm Q}}$ that
is in the friction phase of the string dynamics, then the
vorton density at temperature $\Th$ is given by \cite{Brandenberger:1996zp}
\be \label{26}
\rhv \, = \, {\tilde f} \Bigl({\Th_{_{\rm Q}}\over\mP}\Bigr)^{5/4}
\Bigl({\Th_{_{\rm Q}} \over {\Th_c}}\Bigr)^{3/2} \Th_{_{\rm Q}} \Th^3 \, ,
\ee 
where ${\tilde f}$ is a constant of the (rough) order of $1$. This
result holds both in the radiation and matter dominated phases.

As before, we denote the temperature at which the vortons decay by 
$\Th_{\rm d}$. If the vortons emit a fraction $f$ of their energy as 
photons, then the fractional photon energy density input from vorton 
decay is
\be \label{27}
{{\Delta \rho_\gamma\{\Th_{\rm d}\}} \over {\rho_\gamma\{\Th_{\rm d}\}}} 
\, = \,   \kappa \Bigl({\Th_{_{\rm Q}}\over\mP}\Bigr)^{5/4}
\Bigl({\Th_{_{\rm Q}} \over {\Th_c}}\Bigr)^{3/2} {\Th_{_{\rm Q}} \over 
\Th_{\rm d}} \, ,
\ee
in which  the constant $\kappa$ is given by
\be \label{28}
\kappa \, = \, {{30} \over \pi^2 g_*\{\Th_{\rm d}\} } f {\tilde f} \, ,
\ee
where, $g_*$ denotes the number of spin degrees of freedom in the
radiation bath. The COBE/FIRAS constraint (\ref{fir}) thus
becomes
\be \label{29}
\kappa \Bigl({\Th_{_{\rm Q}}\over\mP}\bigr)^{5/4}
\Bigl({\Th_{_{\rm Q}} \over {\Th_c}}\Bigr)^{3/2} {\Th_{_{\rm Q}} \over
 \Th_{\rm d}} \, < \, 7 \times 10^{-5} \, .
\ee

As a first example, assume that $\Th_{_{\rm Q}} = \Th_c$ (which, 
as shown in (\ref{14}), gives the right order of magnitude for embedded 
strings), and that the decay occurs at the time of last scattering (as again
is reasonable for the embedded strings which are stabilized by
electromagntic plasma effects), i.e. $\Th_{\rm d} = \Th_{\rm ls}$. In this 
case, using the estimate $\kappa \sim 1$, the general constraint (\ref{29})
leads to
\be \label{30}
\Th_c \, \lta \, 10^5 {\rm GeV} \, .
\ee

If we relax the assumption that $\Th_{_{\rm Q}} = \Th_c$ and instead 
parametrize the current condensation time by
\be \label{31}
{\Th_{_{\rm Q}} \over {\Th_c}} \, = \, 10^{\alpha} \, ,
\ee
where $\alpha$ is some (negative) number, the constraint (\ref{29})
becomes
\be \label{32}
\kappa \Bigl({{\Th_c}\over\mP}\Bigr)^{5/4} {{\Th_c} \over {\Th_d}} \, < \, 
7 \times 10^{-5} 10^{-(15/4)\alpha} \, ,
\ee
which corresponds to a severe relaxation of the constraint on $\Th_c$.

\section{Discussion}
\label{sec:10}

We have considered cosmological constraints on models with decaying
topological defects which result from demanding that the photons
produced in the decay do not lead to spectral distortions of the CMB
in excess of the observational limits from the COBE/FIRAS experiment.
The strongest limits arise for theories leading to decaying vortons.
Any theory giving rise to vortons resulting from a string forming phase 
transition above
$10^5$ GeV, and which decayed from a redshift of $10^6$ to today, would
produce a too large spectral distortion as measured by the 
Compton $y$ parameter, and thus be 
ruled out. The above constraint applies both to vortons resulting from
topological strings or those resulting from embedded strings which have
become stabilised by plasma processes. These constraints are
in fact stronger than constraints on vorton models requiring 
compatibility with nucleosynthesis and with the dark matter abundance
limits \cite{Brandenberger:1996zp}.

Since many types of embedded strings are stabilized by interactions
with the electromagnetic plasma, undergo core phase transitions and
become superconducting, thus yielding vortons \cite{Carter:2002te}
which decay at the time of last scattering,
our constraints are very important for theories with embedded defects.

Our analysis also gives rise to constraints on theories with 
non-superconducting topological cosmic strings with GUT scale
symmetry breaking scale, provided they decay in the relevant
redshift interval.

\bigskip
{\bf Acknowledgements}
\medskip

Two of us (ACD and RB) wish to thank W. Unruh and A. Zhitnitsky for 
hospitality at U.B.C., Vancouver, where this work was initiated. 
This work was supported in part by the ESF COSLAB 
programme and by a Royal Society-CNRS exchange grant (BC,ACD), by PPARC (ACD),
by the US Department of Energy under Contract DE-FG0291ER40688, Task A (RB), 
and by an Accord between CNRS and Brown University (BC,RB). We are
grateful to Herb Fried for securing this Accord.


\begin{thebibliography}{}

%\cite{Nagasawa:1999iv}
\bibitem{Nagasawa:1999iv}
M.~Nagasawa and R.~H.~Brandenberger,
%``Stabilization of embedded defects by plasma effects,''
{\it Phys.\ Lett.}\ B {\bf 467}, 205 (1999)
[arXiv:hep-ph/9904261].
%%CITATION = HEP-PH 9904261;%%

%\cite{Vachaspati:dz}
\bibitem{Vachaspati:dz}
T.~Vachaspati and A.~Achucarro,
%``Semilocal Cosmic Strings,''
{\it Phys.\ Rev.}\ D {\bf 44}, 3067 (1991).
%%CITATION = PHRVA,D44,3067;%%

%\cite{Zhang:1997is}
\bibitem{Zhang:1997is}
X.~Zhang, T.~Huang and R.~H.~Brandenberger,
%``Pion and eta strings,''
{\it Phys.\ Rev.}\ D {\bf 58}, 027702 (1998)
[arXiv:hep-ph/9711452].
%%CITATION = HEP-PH 9711452;%%


\bibitem{FIRAS} D. Fixsen et al.,
{\it Astrophys. J.} {\bf 473}, 576 (1996)
[arXiv:astro-ph/9605054].

%\cite{Davis:ij}
\bibitem{Davis:ij}
R.~L.~Davis and E.~P.~Shellard,
%``Cosmic Vortons,''
{\it Nucl.\ Phys.}\ B {\bf 323}, 209 (1989).
%%CITATION = NUPHA,B323,209;%%

%\cite{Brandenberger:1996zp}
\bibitem{Brandenberger:1996zp}
R.~H.~Brandenberger, B.~Carter, A.~C.~Davis and M.~Trodden,
%``Cosmic vortons and particle physics constraints,''
{\it Phys.\ Rev.}\ D {\bf 54}, 6059 (1996)
[arXiv:hep-ph/9605382].
%%CITATION = HEP-PH 9605382;%%


\bibitem{CarterDavis:99} 
B.~Carter and A.~C.~Davis,
%``Chiral vortons and cosmological constraints on particle physics''
{\it Phys.\ Rev.}\ D {\bf 61}, 123501 (2000)
[arXiv:hep-ph/9910560]


%\cite{Carter:2002te}
\bibitem{Carter:2002te}
B.~Carter, R.~H.~Brandenberger and A.~C.~Davis,
``Thermal stabilisation of superconducting sigma strings and their drum  vortons,''
arXiv:hep-ph/0201155.
%%CITATION = HEP-PH 0201155;%%

%\cite{Jeannerot:1999yn}
\bibitem{Jeannerot:1999yn}
R.~Jeannerot, X.~Zhang and R.~H.~Brandenberger,
%``Non-thermal production of neutralino cold dark matter from cosmic  string decays,''
{\it J.H.E.P.} {\bf 9912}, 003 (1999)
[arXiv:hep-ph/9901357].
%%CITATION = HEP-PH 9901357;%%

%\cite{Axenides:1997ja}
\bibitem{Axenides:1997ja}
M.~Axenides and L.~Perivolaropoulos,
%``Topological defects with non-symmetric core,''
Phys.\ Rev.\ D {\bf 56}, 1973 (1997)
[arXiv:hep-ph/9702221].
%%CITATION = HEP-PH 9702221;%%

%\cite{Axenides:1997sk}
\bibitem{Axenides:1997sk}
M.~Axenides, L.~Perivolaropoulos and M.~Trodden,
%``Phase transitions in the core of global embedded defects,''
{\it Phys.\ Rev.} \ D {\bf 58}, 083505 (1998)
[arXiv:hep-ph/9801232].
%%CITATION = HEP-PH 9801232;%%

%\cite{Axenides:1998yz}
\bibitem{Axenides:1998yz}
M.~Axenides, L.~Perivolaropoulos and T.~N.~Tomaras,
%``Core phase structure of cosmic strings and monopoles,''
Phys.\ Rev.\ D {\bf 58}, 103512 (1998)
[arXiv:hep-ph/9803355].
%%CITATION = HEP-PH 9803355;%%

\bibitem{ShellVil}
A. Vilenkin and E.P.S. Shellard, {\it Cosmic strings and other topological 
defects} (Cambridge Univ. Press, Cambridge, 1994).

%\cite{Kibble:1981gv}
\bibitem{Kibble:1981gv}
T.~W.~Kibble,
%``Phase Transitions In The Early Universe,''
Acta Phys.\ Polon.\ B {\bf 13}, 723 (1982).
%%CITATION = APPOA,B13,723;%%

%\cite{Albrecht:2000hu}
\bibitem{Albrecht:2000hu}
A.~Albrecht,
``Defect models of cosmic structure in light of the new CMB data,''
in {\it XXXVth Rencontres de Moriond: Energy Densities in the
       Universe} (2000)
[arXiv:astro-ph/0009129].
%%CITATION = ASTRO-PH 0009129;%%

%\cite{Durrer:2001cg}
\bibitem{Durrer:2001cg}
R.~Durrer, M.~Kunz and A.~Melchiorri,
``Cosmic structure formation with topological defects,''
arXiv:astro-ph/0110348.
%%CITATION = ASTRO-PH 0110348;%%

%\cite{Magueijo:2000se}
\bibitem{Magueijo:2000se}
J.~Magueijo and R.~H.~Brandenberger,
``Cosmic defects and cosmology,''
arXiv:astro-ph/0002030.
%%CITATION = ASTRO-PH 0002030;%%

\bibitem{DimopoulosDavis:98} K.~Dimopoulos and A.~C.~Davis,
%``Friction domination with superconducting strings''
{\it Phys.\ Rev.}\ D {\bf57}, 692-701 (1998)
[arXiv:hep-ph/9705302]

\bibitem{DimopoulosDavis:99} K.~Dimopoulos and A.~C.~Davis,
%``Cosmological consequences of superconducting string networks''
{\it Phys.\ Lett.}\ B {\bf 446}, 236-246 (1999)
[arXiv:hep-ph/9901250]

%\cite{Carter:2000fv}
\bibitem{Carter:2000fv}
B.~Carter, R.~H.~Brandenberger, A.~C.~Davis and G.~Sigl,
%``Prolongation of friction dominated evolution for superconducting cosmic  strings,''
JHEP {\bf 0011}, 019 (2000)
[arXiv:hep-ph/0009278].
%%CITATION = HEP-PH 0009278;%%

%\cite{Davis:1996xs}
\bibitem{Davis:1996xs}
A.~C.~Davis and W.~B.~Perkins,
%``Generic current-carrying strings,''
{\it Phys.\ Lett.}\ B {\bf 390}, 107 (1997)
[arXiv:hep-ph/9610292].
%%CITATION = HEP-PH 9610292;%%



\end{thebibliography}
\end{document}